\documentclass{article}
\usepackage{amssymb}
\usepackage{graphics}
\usepackage{pstricks,psfrag}
\usepackage{amsmath}
\usepackage{subfigure}
\usepackage{vmargin}
\usepackage{epsfig}

\begin{document}

\begin{titlepage}

\title{\bf A Highly Predictive Ansatz for Leptonic Mixing and CP Violation}

\vskip 1cm

\author{G.C. Branco \footnote{e-mail: {\tt gbranco@ist.utl.pt}} and 
J.I. Silva-Marcos \footnote{e-mail: {\tt juca@cftp.ist.utl.pt}} \\ \\
{\it  CFTP, Departamento de F\'{\i}sica} \\
{\it  Instituto Superior T\'ecnico,
Avenida Rovisco Pais, 1} \\
{\it 1049-001 Lisboa, Portugal}}
\maketitle

\vskip 1cm

\begin{abstract}
We suggest a simple highly predictive ansatz 
for charged lepton and light neutrino mass matrices, based on the assumption of 
universality of Yukawa couplings. Using as input the charged lepton masses and light neutrino masses, 
the six parameters characterizing the leptonic mixing matrix $V_{PMNS}$, are predicted in terms of a 
single phase $\phi$, which takes a value around $\phi={\frac{\pi}{2}}$.
Correlations among variuos physical quantities are obtained, in particular $V^{PMNS}_{13}$ is predicted 
as a function of ${\Delta}m^2_{21}$, ${\Delta}m^2_{31}$ and 
$\sin^2(\theta_{sol})$, and restricted to the range 
$0.167<|V^{PMNS}_{13}|<0.179$. 
\end{abstract}
\vskip 1cm
PACS numbers :~12.10.Kt, 12.15.Ff, 14.60.Pq

\end{titlepage}

\section{\protect\bigskip Introduction}

\bigskip Understanding the pattern of fermion masses and flavour mixing is
still one of the open fundamental questions in particle physics. The
discovery of large leptonic mixing, in contrast to small quark mixing, has
rendered the flavour puzzle even more intriguing.

In the Standard Model (SM) and in most of its extensions, the arbitrariness
of fermion masses and mixing stems from the fact that the gauge invariance
does not constrain the flavour structure of the Yukawa couplings. The fact
that, in the SM, only Yukawa couplings can be complex, has motivated the
hypothesis of universality of strength of Yukawa couplings (USY) \cite%
{usy-first}, which would all have the same strength, with flavour-dependent
phases. The consequences of USY have been analyzed in various works, both
for the quark \cite{usy-second} and lepton sectors \cite{akhmedov}. Such an
USY structure for the Yukawa couplings could arise from higher-dimensional
theories \cite{hung-seco}.

In this paper, we suggest a highly predictive USY ansatz which is able to
accommodate our present experimental knowledge on lepton masses and mixing
and makes definite predictions, which can be tested in the near future. More
specifically, in this USY ansatz, once the charged lepton and neutrino
masses are fixed, the three leptonic mixing angles, the Dirac phases and the
two Majorana phases are all predicted in terms of only one free parameter.
This highly constrained system implies interesting correlations among
various physical quantities.

The size of $V_{13}^{PMNS}$ is predicted as a function of $\tan (\theta
_{sol})$ and the neutrino mass differences $\Delta m_{21}^{2}$, $\Delta
m_{31}^{2}$. For central values of $\sin ^{2}(\theta _{sol})$ and $\Delta
m_{ij}^{2}$, one obtains $\left\vert V_{13}^{PMNS}\right\vert =0.178$,
clearly at the reach of the next round of experiments \cite{kayser}. The
ansatz also predicts the strength of Dirac type CP violation, measured by
the invariant quartet $I_{CP}\equiv \left\vert \mathrm{Im}\left[ V_{12}\
V_{23}\ V_{22}^{\ast }V_{13}^{\ast }\right] \right\vert ^{PMNS}$ . For
central values of $\sin ^{2}(\theta _{atm})$, $\sin ^{2}(\theta _{sol})$ and 
$\Delta m_{ij}^{2}$, one obtains $I_{CP}=0.00906$ , which can be measured in
neutrino oscillation experiments \cite{kayser}.

This paper is organized as follows. In the next section, we describe the
Ansatz and its parameter space, both in the charged lepton and neutrino
sectors. In section 3, we evaluate the lepton mixing and derive some
predictions of the ansatz for various physical quantities, including $%
\left\vert V_{13}^{PMNS}\right\vert $, double beta decay and the strength of
the Dirac-type CP violation. Section 4 contains some numerical results and
figures illustrating correlations among various physical quantities.
Finally, our conclusions are contained in section 5.

\section{The Ansatz and its Parameter Space}

\subsection{The charged lepton sector}

We propose the following USY structure for the charged lepton mass matrix%
\begin{equation}
M_{l}=\frac{c_{l}}{\sqrt{3}}\cdot K_{\phi }^{\dagger }\cdot \left[ 
\begin{array}{ccc}
1 & 1 & 1 \\ 
1 & e^{ia_{l}} & 1 \\ 
1 & 1 & e^{ib_{l}}%
\end{array}%
\right] \qquad ;\qquad K_{\phi }=diag(1,1,e^{i\phi })  \label{ml}
\end{equation}%
The phase $\phi $\ does not affect the charged lepton mass spectrum but
contributes to the leptonic mixing. Using the trace, determinant and second
invariant of $H_{l}\equiv M_{l}M_{l}^{\dagger }$, one can derive exact
expressions for the phases $a_{l},b_{l}$ and the parameter $c_{l}$ in terms
of the masses: 
\begin{equation}
\begin{array}{l}
c_{l}=\frac{1}{\sqrt{3}}\sqrt{m_{\tau }^{2}+m_{\mu }^{2}+m_{e}^{2}} \\ 
\\ 
3\sin ^{2}(\frac{a_{l}}{2})+3\sin ^{2}(\frac{b_{l}}{2})+\sin ^{2}(\frac{%
a_{l}+b_{l}}{2})=\frac{81}{4}\frac{m_{\mu }^{2}m_{\tau
}^{2}+m_{e}^{2}m_{\tau }^{2}+m_{e}^{2}m_{\mu }^{2}}{\left( m_{\tau
}^{2}+m_{\mu }^{2}+m_{e}^{2}\right) ^{2}} \\ 
\\ 
\left\vert \sin (\frac{a_{l}}{2})\sin (\frac{b_{l}}{2})\right\vert =\frac{27%
}{4}\frac{m_{e}m_{\mu }m_{\tau }}{\sqrt{\left( m_{\tau }^{2}+m_{\mu
}^{2}+m_{e}^{2}\right) ^{3}}}%
\end{array}
\label{dechi-lep}
\end{equation}%
From the charged lepton hierarchy, one obtains to an excellent approximation%
\begin{equation}
|a_{l}|\simeq 6\frac{m_{e}}{m_{\tau }}\,,~\quad \quad |b_{l}|\simeq \frac{9}{%
2}\frac{m_{\mu }}{m_{\tau }}\,,  \label{albl}
\end{equation}%
Obviously, in Eq.(\ref{dechi-lep}) $a_{l}$ and $b_{l}$ enter in a symmetric
way. The choice of Eq.(\ref{albl}) is required in order to obtain the right
eigenvalue ordering.

\subsection{The effective neutrino mass matrix}

We assume that lepton number is violated at a high energy scale, leading at
low energies to the following effective neutrino mass matrix, 
\begin{equation}
M_{\nu }=\frac{c_{\nu }}{\sqrt{3}}\left[ 
\begin{array}{ccc}
e^{ia_{\nu }} & 1 & 1 \\ 
1 & e^{-ia_{\nu }} & 1 \\ 
1 & 1 & e^{ib_{\nu }}%
\end{array}%
\right]   \label{neut3}
\end{equation}%
The three parameters, $c_{\nu },b_{\nu }$ and $a_{\nu }$ of the neutrino
mass matrix ansatz in Eq. (\ref{neut3}) are entirely determined by the three
neutrino masses. We find from the trace, second invariant and determinant of 
$H_{\nu }\equiv M_{\nu }M_{\nu }^{\dagger }$:%
\begin{equation}
\begin{array}{l}
3c_{\nu }^{2}=m_{3}^{2}+m_{2}^{2}+m_{1}^{2} \\ 
\\ 
\cos (a_{\nu })=1-\frac{27}{2}d_{\nu } \\ 
\\ 
\cos (b_{\nu })=\frac{1+\frac{27}{2}d_{\nu }-\frac{81}{8}\chi _{\nu }}{1-%
\frac{27}{4}d_{\nu }}%
\end{array}
\label{angleso}
\end{equation}%
where%
\begin{equation}
d_{\nu }=\frac{m_{1}m_{2}m_{3}}{\left( m_{3}^{2}+m_{2}^{2}+m_{1}^{2}\right)
^{\frac{3}{2}}}\qquad ;\qquad \chi _{\nu }=\frac{%
m_{1}^{2}m_{2}^{2}+m_{1}^{2}m_{3}^{2}+m_{2}^{2}m_{3}^{2}}{\left(
m_{3}^{2}+m_{2}^{2}+m_{1}^{2}\right) ^{2}}  \label{angles}
\end{equation}

At this stage, it is worth emphasizing the predictive power of the Ansatz.
From Eqs. (\ref{dechi-lep}, \ref{angleso}), it is clear that once the
parameters $(c_{l},b_{l}$, $a_{l})$, and $(c_{\nu }$, $b_{\nu }$, $a_{\nu })$
are fixed by the charged lepton and neutrino masses, the six parameters of
the Pontecorvo--Maki--Nakagawa--Sakata matrix, $V_{PMNS}$, are completely
determined in terms of a single parameter, the phase $\phi $.

\section{Evaluation of Lepton Mixing}

\subsection{Diagonalization and parametrization of the lepton mass matrices}

The diagonalization of the Hermitian charged lepton mass matrix $H_{l}\equiv
M_{l}M_{l}^{\dagger }$ is carried out through%
\begin{equation*}
V_{l}^{\dagger }\cdot H_{l}\cdot V_{l}=diag(m_{e}^{2},m_{\mu }^{2},m_{\tau
}^{2})
\end{equation*}%
with the unitary matrix $V_{l}$\ given by 
\begin{equation*}
V_{l}=K_{\phi }^{\dagger }\cdot F\cdot W_{l}
\end{equation*}%
where $F$ 
\begin{equation}
F=\left( 
\begin{array}{ccc}
~~\frac{1}{\sqrt{2}} & ~~\frac{-1}{\sqrt{6}} & ~~\frac{1}{\sqrt{3}} \\ 
-\frac{1}{\sqrt{2}} & ~~\frac{-1}{\sqrt{6}} & ~~\frac{1}{\sqrt{3}} \\ 
~0 & \frac{2}{\sqrt{6}} & ~~\frac{1}{\sqrt{3}}%
\end{array}%
\right) \,  \label{f}
\end{equation}%
and $W_{l}$ is a unitary matrix close to the identity. Given the strong
hierarchy of the charged lepton masses, to an excellent approximation, one
obtains for $W_{l}$ 
\begin{equation}
W_{l}\simeq {\left( 
\begin{array}{ccc}
\vspace*{0.15cm}1 & \frac{m_{e}}{\sqrt{3}m_{\mu }} & -i\sqrt{\frac{2}{3}}%
\frac{m_{e}}{m_{\tau }} \\ 
\vspace*{0.15cm}-\frac{m_{e}}{\sqrt{3}m_{\mu }} & 1-\frac{1}{2}\left( \frac{%
m_{\mu }}{m_{\tau }}\right) ^{2} & i\frac{m_{\mu }}{\sqrt{2}m_{\tau }} \\ 
-i\sqrt{\frac{3}{2}}\frac{m_{e}}{m_{\tau }} & i\frac{m_{\mu }}{\sqrt{2}%
m_{\tau }} & 1-\frac{1}{2}\left( \frac{m_{\mu }}{m_{\tau }}\right) ^{2}%
\end{array}%
\right) }  \label{wl}
\end{equation}

The diagonalization of the neutrino mass matrix is achieved through%
\begin{equation}
V_{\nu }^{\dagger }\cdot M_{\nu }\cdot V_{\nu }^{\ast }=D_{\nu
}=diag(m_{1},m_{2},m_{3})  \label{diagmv}
\end{equation}%
where $m_{i}$ denote the neutrino masses. In order to understand the main
features of $V_{\nu }$ in the framework of our Ansatz, it is useful to
introduce a convenient parametrization. Let us now introduce the
dimensionless parameters $\varepsilon $, $\delta $ defined by%
\begin{equation}
\varepsilon ~=\frac{m_{_{2}}}{\sqrt{m_{3}^{2}+m_{2}^{2}+m_{1}^{2}}}\qquad
;\qquad \delta =\frac{m_{1}}{m_{2}}  \label{ep-delta-1}
\end{equation}%
The neutrino masses can then be written:

\begin{equation}
\begin{array}{l}
m_{1}=\sqrt{3}\ c_{\nu }~\varepsilon ~\delta ~ \\ 
m_{2}=\sqrt{3}\ c_{\nu }~\varepsilon ~ \\ 
m_{3}=\sqrt{3}\ c_{\nu }\sqrt{1-\varepsilon ^{2}-\delta ^{2}\varepsilon ^{2}}%
\end{array}
\label{ep-delta-2}
\end{equation}

By substituting $m_{i}$ as functions of $\varepsilon$, $\delta $ in Eqs. (%
\ref{angleso}, \ref{angles}), we obtain $d_{\nu },\chi _{\nu }$ as well as $%
a_{\nu },b_{\nu }$ as functions of $\varepsilon$ and $\delta $: 
\begin{equation}
d_{\nu }=\delta \ \varepsilon ^{2}\ \sqrt{1-\varepsilon ^{2}(1+\delta ^{2})}%
\ ;\qquad \chi _{\nu }=\varepsilon ^{2}\ \left[ 1+\delta ^{2}-\varepsilon
^{2}\ (1+\delta ^{2}+\delta ^{4})\right]  \label{angles1}
\end{equation}%
The matrix $V_{\nu }$ is then entirely given as a function of these two
parameters ($\varepsilon$, $\delta $), which are fixed by neutrino mass
ratios. Furthermore, for our ansatz, $V_{\nu }$ is exactly factorizable in
the following way: 
\begin{equation}
V_{\nu }=F\cdot K_{\gamma }\cdot O_{\nu }\cdot K_{M}  \label{vn}
\end{equation}%
where $F$ was given in Eq. (\ref{f}) and $K_{\gamma }$, $K_{M}$ are diagonal
unitary matrices containing phases, which will contribute to the Dirac and
Majorana type phases of the lepton mixing, $K_{\gamma }=$ $diag(1,e^{i\gamma
},-i)$ and $K_{M}=diag(e^{i\widehat{\alpha }_{M}},e^{i\widehat{\beta }%
_{M}},e^{i\widehat{\gamma }_{M}})$. As mentioned, all these phases and the
angles of orthogonal matrix $O_{\nu }$ can be expressed as functions of $%
\delta $ and $\varepsilon $.

So far all our results are exact. Our numerical results for $V_{PMNS}$ will
be obtained through exact numerical diagonalization of $H_{l}$ and $H_{\nu }$%
. However, in order to get an overview of the physical implications of this
USY ansatz, it is useful to derive some analytical expressions which hold to
a good approximation. \ Let us parametrize $O_{\nu }$\ in the following way: 
\begin{equation}
O_{\nu }=O_{23}\cdot O_{13}\cdot O_{12}  \label{orth}
\end{equation}%
with%
\begin{equation*}
\begin{array}{l}
O_{23}=\left[ 
\begin{array}{ccc}
1 & 0 & 0 \\ 
0 & \cos (\widehat{\theta }_{23}) & \sin (\widehat{\theta }_{23}) \\ 
0 & -\sin (\widehat{\theta }_{23}) & \cos (\widehat{\theta }_{23})%
\end{array}%
\right] ;\qquad O_{13}=\left[ 
\begin{array}{ccc}
\cos (\widehat{\theta }_{13}) & 0 & \sin (\widehat{\theta }_{13}) \\ 
0 & 1 & 0 \\ 
-\sin (\widehat{\theta }_{13}) & 0 & \cos (\widehat{\theta }_{13})%
\end{array}%
\right] \\ 
\\ 
O_{12}=\left[ 
\begin{array}{ccc}
\cos (\widehat{\theta }_{12}) & \sin (\widehat{\theta }_{12}) & 0 \\ 
-\sin (\widehat{\theta }_{12}) & \cos (\widehat{\theta }_{12}) & 0 \\ 
0 & 0 & 1%
\end{array}%
\right]%
\end{array}%
\end{equation*}

It turns out that in the relevant region of parameter space, $\varepsilon $
is relatively small, $\varepsilon \approx 0.2$. Therefore, we make an
expansion in powers of $\varepsilon $ which yield: 
\begin{equation}
\begin{array}{l}
\tan (\widehat{\theta }_{12})=-\sqrt{\delta }\left( 1+\frac{8\delta +8\delta
^{2}-3\delta ^{3}-3}{4(1-\delta )}\varepsilon ^{2}+O(\varepsilon ^{4})\right)
\\ 
\\ 
\tan (\widehat{\theta }_{23})=\varepsilon \ \frac{(1-\delta )}{\sqrt{2}}%
\left( 1+\frac{37\delta -2\delta ^{2}-2}{8}\varepsilon ^{2}+O(\varepsilon
^{4})\right) \\ 
\\ 
\tan (\widehat{\theta }_{13})=\varepsilon \ \sqrt{2\delta }\left( 1+\frac{%
4+\delta +4\delta ^{2}}{8}\varepsilon ^{2}+O(\varepsilon ^{4})\right) \\ 
\\ 
\tan (\gamma )=-\varepsilon \ \frac{(10\delta -3\delta ^{2}-3)}{4(1-\delta )}%
+O(\varepsilon ^{3})%
\end{array}
\label{wmatrix}
\end{equation}

The leptonic mixing matrix is given by:%
\begin{equation}
V_{PMNS}=V_{l}^{\dagger }\cdot V_{\nu }=\left( W_{l}^{\dagger }\ F^{T}\
K_{\phi }\right) \cdot \left( F\ K_{\gamma }\ O_{\nu }\ K_{M}\right) 
\label{vpmns}
\end{equation}%
This formula is exact and it will be used in the numerical computation of $%
V_{PMNS}$. However, it is useful to obtain analytical approximate
expressions for $V_{PMNS}$. Using Eqs. (\ref{vpmns}, \ref{orth}), and
neglecting the small contribution from $W_{l}$ given by Eq. (\ref{wl}), one
obtains 
\begin{equation}
\begin{array}{l}
\left\vert \frac{V_{12}^{PMNS}}{V_{11}^{PMNS}}\right\vert \equiv \left\vert
\tan (\theta _{sol})\right\vert =\left\vert \tan (\widehat{\theta }%
_{12})\right\vert  \\ 
\\ 
|V_{13}^{PMNS}|=|\sin (\widehat{\theta }_{13})|%
\end{array}
\label{v12v11}
\end{equation}%
which identifies these two lepton mixing angles in terms of our
parametrization. Up to second order in $\varepsilon $, from Eq. (\ref%
{wmatrix}), one obtains $\tan ^{2}(\theta _{sol})$ and $|V_{13}^{PMNS}|$
expressed in terms of the measured $\Delta m_{31}^{2}$, $\Delta m_{21}^{2}$
and the lightest neutrino mass, $m_{1}$: 
\begin{equation}
\begin{array}{l}
\tan ^{2}(\theta _{sol})=\frac{m_{1}}{\sqrt{\Delta m_{21}^{2}+m_{1}^{2}}} \\ 
\\ 
\left\vert V_{13}^{PMNS}\right\vert ^{2}=\frac{2m_{1}\sqrt{\Delta
m_{21}^{2}+m_{1}^{2}}}{\Delta m_{31}^{2}+\Delta m_{21}^{2}+3m_{1}^{2}}%
\end{array}
\label{relde-1}
\end{equation}%
Eliminating $m_{1}$ from Eq. (\ref{relde-1}), one obtains the interesting
sum rule expressing $|V_{13}^{PMNS}|$ in terms of measured quantities%
\begin{equation}
\left\vert V_{13}^{PMNS}\right\vert =\sqrt{2}\left\vert \tan (\theta
_{sol})\right\vert \sqrt{\frac{\Delta m_{21}^{2}}{\Delta m_{31}^{2}}}~\frac{1%
}{\sqrt{1-\tan ^{4}(\theta _{sol})+\left( 1+2\tan ^{4}(\theta _{sol})\right) 
\frac{\Delta m_{21}^{2}}{\Delta m_{31}^{2}}}}  \label{relde-b}
\end{equation}%
For central values of $\sin ^{2}(\theta _{sol})$ \ and $\Delta m_{ij}^{2}$
one finds 
\begin{equation}
\left\vert V_{13}^{PMNS}\right\vert =0.178
\end{equation}%
For $\theta _{atm}$ one obtains: 
\begin{equation}
\sin ^{2}(\theta _{atm})=\frac{4}{9}\left[ 1-\cos (\phi )+\frac{3}{2}\
\varepsilon \ (1-\delta )\sin (\phi )+O(\varepsilon ^{2})\right] 
\label{atmo}
\end{equation}%
It is clear that $\theta _{atm}$ crucially depends on $\phi $, the phase
defined in Eq.(\ref{ml}). It is interesting to note that a good fit of $%
\theta _{atm}$\ is obtained for $\phi =\frac{\pi }{2}$.

\subsection{Double Beta Decay}

We evaluate now $M_{ee}$, which controls the strength of double beta decay
and is given by:

\begin{equation}
M_{ee}\equiv \left\vert m_{1}\ \left( V_{11}^{PMNS}\right) ^{2}+m_{2}\
\left( V_{12}^{PMNS}\right) ^{2}+m_{3}\ \left( V_{13}^{PMNS}\right)
^{2}\right\vert  \label{mee}
\end{equation}%
We compute $M_{ee}$ in two steps. First, we evaluate the contribution to
Majorana phases from $K_{M}=diag(e^{i\widehat{\alpha }_{M}},e^{i\widehat{%
\beta }_{M}},e^{i\widehat{\gamma }_{M}})$. This can be done by focussing
only on the diagonalization of the neutrino mass matrix: $V_{\nu }^{\dagger
}\cdot M_{\nu }\cdot $ $V_{\nu }^{\ast }=diag(m_{1},m_{2},m_{3})$. It is
clear that these phases appear when diagonalizing $M_{\nu }$ only with $F\
K_{\gamma }\ O_{\nu }$, without $K_{M}$ : 
\begin{equation}
\left( F\ K_{\gamma }\ O_{\nu }\right) ^{\dagger }\cdot M_{\nu }\cdot \left(
F\ K_{\gamma }\ O_{\nu }\right) ^{\ast }=diag(m_{1}\ e^{2i\widehat{\alpha }%
_{M}},m_{2}\ e^{2i\widehat{\beta }_{M}},m_{3}\ e^{2i\widehat{\gamma }_{M}})
\label{mass-maj}
\end{equation}%
In leading order, we find\footnote{%
Obviously, the phases $\widehat{\alpha }_{M}$, $\widehat{\beta }_{M}$, $%
\widehat{\gamma }_{M}$ are defined modulo $\pi $.}%
\begin{equation}
\begin{array}{l}
2\widehat{\alpha }_{M}=-\frac{\pi }{2}-\frac{9-12\delta -\delta ^{2}}{%
4(1-\delta )}\varepsilon \\ 
2\widehat{\beta }_{M}=\frac{\pi }{2}+\frac{1+12\delta -9\delta ^{2}}{%
4(1-\delta )}\varepsilon \\ 
2\widehat{\gamma }_{M}=\pi +\frac{(1-\delta )}{2}\varepsilon%
\end{array}
\label{maj}
\end{equation}%
We can then write \ 
\begin{equation}
M_{ee}=\left\vert m_{1}\ e^{2i\widehat{\alpha }_{M}}\left( V_{11}\right)
^{2}+m_{2}\ e^{2i\widehat{\beta }_{M}}\left( V_{12}\right) ^{2}+m_{3}\ e^{2i%
\widehat{\gamma }_{M}}\left( V_{13}\right) ^{2}\right\vert  \label{mee-a}
\end{equation}%
where here $V$ is the lepton mixing matrix $V_{PMNS}$ but without the last $%
K_{M}$ phases, i.e. $V=V_{PMNS}\cdot K_{M}^{\ast }=W_{l}^{\dagger }\ F^{T}\
K_{\phi }\ F\ K_{\gamma }\ O_{\nu }$.

Since,\ the matrix $F^{T}\ K_{\phi }\ F$ in $V$ only gives a contribution in
the 2--3 plane, and $W^{l}_{12}$ and $W^{l}_{13}$ are all of the order of $%
\varepsilon ^{5}$ or smaller, we may read the expressions for $V_{11}$, $%
V_{12}$ and $V_{13}$ directly from the leading order expressions for 
\begin{equation}
\begin{array}{l}
\tan (\widehat{\theta }_{12})=-\sqrt{\delta }\ \left( 1+\frac{8\delta
+8\delta ^{2}-3\delta ^{3}-3}{4(1-\delta )}\varepsilon ^{2}\right) \\ 
\\ 
\tan (\widehat{\theta }_{13})=\varepsilon \ \sqrt{2\delta }\ \left( 1+\frac{%
4+\delta +4\delta ^{2}}{8}\varepsilon ^{2}\right) \\ 
\\ 
\tan (\gamma )=-\varepsilon \ \frac{(10\delta -3\delta ^{2}-3)}{4(1-\delta )}%
\end{array}
\label{wmatrix-a}
\end{equation}%
Using Eqs. (\ref{maj}, \ref{mee-a}, \ref{wmatrix-a}) together with $m_{1}$, $%
m_{2}$ and $m_{3}$ expressed in terms of $c_{\nu}$, $\delta$ and $\varepsilon
$ (as in Eq. (\ref{ep-delta-2})), we find the following leading order
expression 
\begin{equation}
M_{ee}=\frac{9\sqrt{3}}{2}\delta \varepsilon ^{2}\left( 1-\frac{(1+\delta
^{2})}{2}\varepsilon ^{2}-\frac{(1+\delta ^{2})^{2}}{8}\varepsilon
^{4}\right) \ c_{\nu }  \label{mee-b}
\end{equation}

\subsection{\protect\bigskip Dirac-type CP Violation}

The strength of the Dirac-type CP violation is given by the imaginary part
of any rephasing invariant quartet of $V_{PMNS}$, e.g. 
\begin{equation}
I_{CP}=\left\vert \mathrm{{Im}\left[ V_{12}\ V_{23}\ V_{22}^{\ast
}V_{13}^{\ast }\right] }^{PMNS}\right\vert   \label{jcp}
\end{equation}%
Using Eqs. (\ref{ml}, \ref{f}, \ref{wl}, \ref{orth}), we can evaluate $I_{CP}
$ in terms of $\varepsilon $, $\delta $ and $\phi $, obtaining in second
order of $\varepsilon :$ 
\begin{equation}
I_{CP}=\frac{2\delta \varepsilon }{9(1+\delta )}\left[ 1-\cos (\phi
)-\varepsilon \ \frac{3(10\delta -3\delta ^{2}-3)}{4(1-\delta )}\sin (\phi )%
\right]   \label{jcpab}
\end{equation}%
From Eq. ( \ref{atmo}), it is clear that the phase $\phi $ is strongly
correlated with $\sin (\theta _{atm})$. Then, for the central value of $\sin
^{2}(\theta _{atm})$, which is obtained with $\phi =\frac{\pi }{2}$, and
central values of $\sin ^{2}(\theta _{sol})$ and $\Delta m_{ij}^{2}$, one
gets $\left\vert I_{CP}\right\vert =0.0105$, a value obtained neglecting the
charged lepton contribution, which is small. Further on, in Section 4, we
shall give an exact numerical example.

\subsection{Majorana-type CP Violation}

It is well known that in the case of Majorana neutrinos, the basic rephasing
invariants, in the leptonic sector, are bilinears of the type 
$V_{jk}^{PMNS}V_{jl}^{{PMNS}\ast }$ with $k\neq l$. In fact, 
in 
the
case of three leptonic flavours, it has recently been shown that there are
six rephasing invariant independent "Majorana-type" phases from which one
can reconstruct the full $V_{PMNS}$ matrix\ using $3\times 3$ unitarity \cite%
{invariants}. One can choose as basic Majorana phases%
\begin{equation}
\begin{array}{lll}
\gamma _{1}=Arg\left[ V_{11}\left( V_{13}\right) ^{\ast }\right] &  & \beta
_{1}=Arg\left[ V_{12}\left( V_{13}\right) ^{\ast }\right] \\ 
&  &  \\ 
\gamma _{2}=Arg\left[ V_{21}\left( V_{23}\right) ^{\ast }\right] &  & \beta
_{2}=Arg\left[ V_{22}\left( V_{23}\right) ^{\ast }\right] \\ 
&  &  \\ 
\gamma _{3}=Arg\left[ V_{31}\left( V_{33}\right) ^{\ast }\right] &  & \beta
_{3}=Arg\left[ V_{32}\left( V_{33}\right) ^{\ast }\right]%
\end{array}
\label{arg}
\end{equation}%
where we have dropped the PMNS superscript in the $V_{ij}$'s. The $\gamma
_{i}$, $\beta _{i}$ can be evaluated in the present USY ansatz and we obtain
in leading order, 
\begin{equation}
\begin{array}{lll}
\gamma _{1}=-\frac{3\pi }{4}-\frac{11-16\delta +\delta ^{2}}{8(1-\delta )}%
\varepsilon &  & \beta _{1}=-\frac{\pi }{4}+\frac{16\delta -11\delta ^{2}-1}{%
8(1-\delta )}\varepsilon \\ 
&  &  \\ 
\gamma _{2}=-\frac{\pi }{4}-\arctan \left( \frac{3\sin (\phi )}{1-\cos (\phi
)}\right) &  & \beta _{2}=\frac{5\pi }{4}-\arctan \left( \frac{3\sin (\phi )%
}{1-\cos (\phi )}\right) \\ 
&  &  \\ 
\gamma _{3}=\frac{3\pi }{4}-\arctan \left( \frac{3\sin (\phi )}{1-\cos (\phi
)}\right) &  & \beta _{3}=\frac{\pi }{4}-\arctan \left( \frac{3\sin (\phi )}{%
1-\cos (\phi )}\right)%
\end{array}
\label{arg1}
\end{equation}

\bigskip

\section{Numerical Results}

The predictive power of our Ansatz is best shown with a set of figures. Fig.
1 demonstrates the dependence of the solar mixing angle on the value of $%
m_{1}$. We allow for explicit experimental uncertainties of $\Delta
m_{21}^{2}=7.65_{-20}^{+23}\times 10^{-5}\ eV^{2}$ and $\Delta
m_{31}^{2}=2.40_{-11}^{+12}\times 10^{-3}\ eV^{2}$. It is clear, that a
central value for $\sin ^{2}(\theta _{sol})=0.3$ implies a prediction for
the value of $m_{1}$: $0.0316eV<m_{1}<0.0345eV$. From Fig. 2 and 3, it also
follows that $0.167<|V_{13}^{PMNS}|<0.179$ and that $0.00315<M_{ee}<$ $%
0.00345$. Notice that, choosing the neutrino mass differences and $\sin
^{2}(\theta _{sol})=0.304_{-16}^{+22}$ within these $1\sigma $ experimental
constraints, our model accommodates the upperlimit for $%
|V_{13}^{PMNS}|^{2}<0.004$. The mixing angle $\sin ^{2}(\theta _{atm})$ and
the experimental observable measuring CP violation $I_{CP}$ depend crucially
on the angle $\phi $ and thus we may plot the two experimental observables
against each other. From Fig. 4 we find, that for a central value of $\sin
^{2}(\theta _{sol})=0.5$, that $0.0090<I_{CP}<0.0098$.

Next, we give an explicit numerical example, where six of the input
parameters of the Ansatz are fixed by the known charged lepton masses, two
neutrino mass differences $\Delta m_{21}^{2},\Delta m_{31}^{2}$, together
with a chosen value for the lightest neutrino mass $m_{1}$. Then, the six
parameters of $V_{PMNS}$\ are all predicted with a single free parameter,
namely the phase $\phi $, which is taken to be $\phi =\frac{\pi }{2}$.

INPUT:%
\begin{equation*}
\begin{array}{lllll}
c_{l}=1023.72\ \ eV &  & a_{l}=1.729\times 10^{-3} &  & a_{\nu} =0.66 \\ 
&  & b_{l}=0.2677 &  & b_{\nu} =0.5077 \\ 
c_{\nu }=0.0290352\ \ eV &  & \phi =\frac{\pi }{2} &  & 
\end{array}%
\end{equation*}%
where, for this particular example we have $\delta =0.4286$ and $%
\varepsilon=0.1927$. We then find

OUTPUT%
\begin{equation*}
|V_{PMNS}|=\left[ 
\begin{array}{ccc}
0.81573 & 0.55015 & 0.17867 \\ 
0.30173 & 0.66298 & 0.68514 \\ 
0.49350 & 0.50773 & 0.70616%
\end{array}%
\right] \quad ;\quad
\end{equation*}%
with%
\begin{equation*}
\sin ^{2}(\theta _{sol})=0.313\quad ;\quad \sin ^{2}(\theta
_{atm})=0.485\quad ;\quad |V_{13}^{PMNS}|^{2}=0.0319
\end{equation*}%
and%
\begin{equation*}
\begin{array}{lll}
m_{e}=0.51\ MeV & m_{1}=4.15\times 10^{-3}\ eV & \Delta
m_{21}^{2}=7.664\times 10^{-5}\ eV^{2} \\ 
m_{\mu }=105.5\ MeV & m_{2}=9.69\times 10^{-3}\ eV & \Delta
m_{31}^{2}=2.401\times 10^{-3}\ eV^{2} \\ 
m_{\tau }=1770\ MeV & m_{3}=0.04917\ eV\quad & 
\end{array}%
\end{equation*}%
We obtain for the Majorana observables%
\begin{equation*}
Arg\left( 
\begin{array}{cc}
V_{11}V_{13}^{\ast } & V_{12}V_{13}^{\ast } \\ 
V_{21}V_{23}^{\ast } & V_{22}V_{23}^{\ast } \\ 
V_{31}V_{33}^{\ast } & V_{32}V_{33}^{\ast }%
\end{array}%
\right) ^{PMNS}=\left( 
\begin{array}{rr}
-2.535 & -0.6145 \\ 
-2.230 & 2.731 \\ 
0.7857 & -0.3545%
\end{array}%
\right)
\end{equation*}%
and for the strength of the Dirac type CP violation and double beta decay%
\begin{equation*}
M_{ee}=3.53\times 10^{-3}\ eV\quad ;\quad I_{CP}=0.00906
\end{equation*}

\section{Conclusions}

\bigskip We have pointed out that a simple ansatz, inspired by the
hypothesis of universality of Yukawa couplings, leads to a highly predictive
scheme for leptonic mixing. If one uses as input the charged lepton and
neutrino masses, then the three mixing angles and the three CP violating
phases entering in $V_{PMNS}$ are all predicted in terms of a single phase
which takes the value $\phi \approx \frac{\pi }{2}$. The Ansatz predicts a
relatively large value of $|V_{13}^{PMNS}|$ and of $I_{CP}$, clearly at the
reach of the next round of experiments \cite{kayser}.

At this stage, it is worth recalling that when applied to the quark sector,
the USY hypothesis can accommodate the main features of the CKM matrix, but
cannot account for the observed strength of CP violation in the quark
sector, measured by the rephasing invariant $|\mathrm{Im}%
[V_{ub}V_{cb}V_{ub}^{\ast }V_{cs}^{\ast }]|.$However, it has been recently
pointed out \cite{yamamoto} that sufficient CP violation can be obtained in
extensions of the Standard Model where a USY structure is assumed but extra
down singlet quarks are introduced and mix with the standard quarks.

The USY ansatz has clearly a great appeal. A crucial open question is
finding a symmetry principle, eventually implemented in a framework with
extra dimensions \cite{hung-seco}, which can naturally lead to the
universality of the strength of Yukawa couplings.

\section*{Acknowledgements}

This work was partially supported by Funda\c c\~ao para a Ci\^encia e a
Tecnologia (FCT, Portugal) through the Projects POCI/81919/2007, CFTP-FCT
UNIT 777, and by CERN/FP/83502/2008, which are partially funded through
POCTI (FEDER), and by the Marie Curie RTNs MRT-CT-2006-035505 and
MRT-CT-503369.

\begin{figure}[tbp]
\centerline{\ \epsfysize=8.0truecm \epsffile{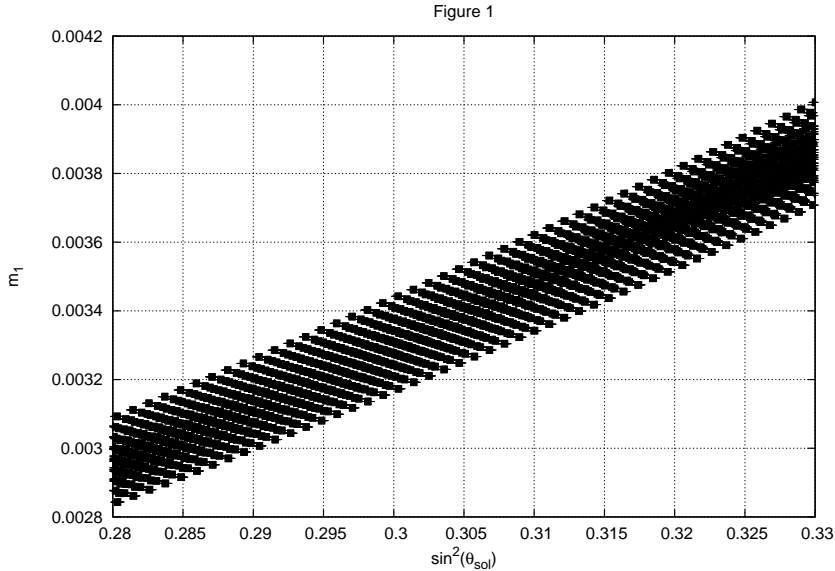}}
\caption{$m_{1}$ as a function of $\sin ^{2}{\protect\theta _{sol}}$,
assuming $1\protect\sigma $ uncertainties in neutrino mass differencies}
\label{fig1}
\end{figure}

\begin{figure}[tbp]
\centerline{\ \epsfysize=8.0truecm \epsffile{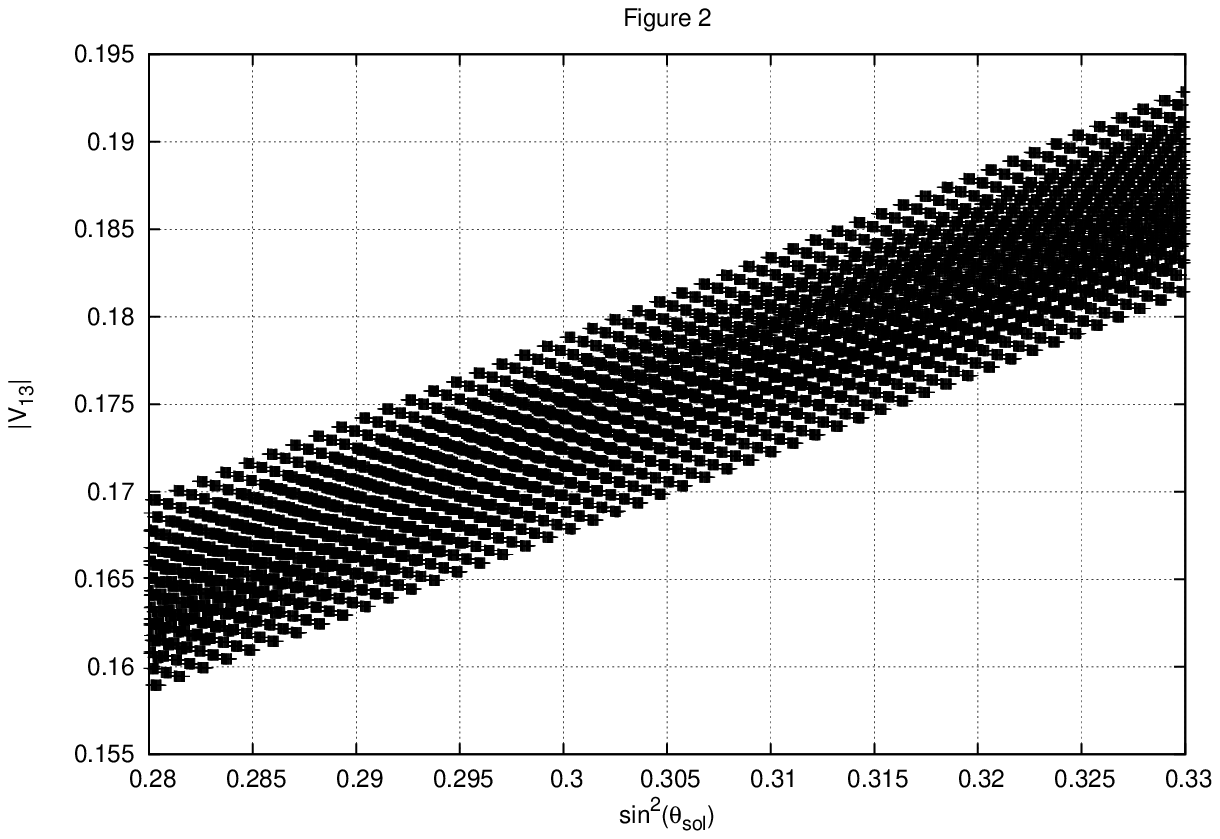}}
\caption{$|V_{13}| $ as a function of $\sin^{2}{\protect\theta_{sol}}$,
assuming $1\protect\sigma$ uncertainties in neutrino mass differencies}
\label{fig2}
\end{figure}

\begin{figure}[tbp]
\centerline{\ \epsfysize=8.0truecm \epsffile{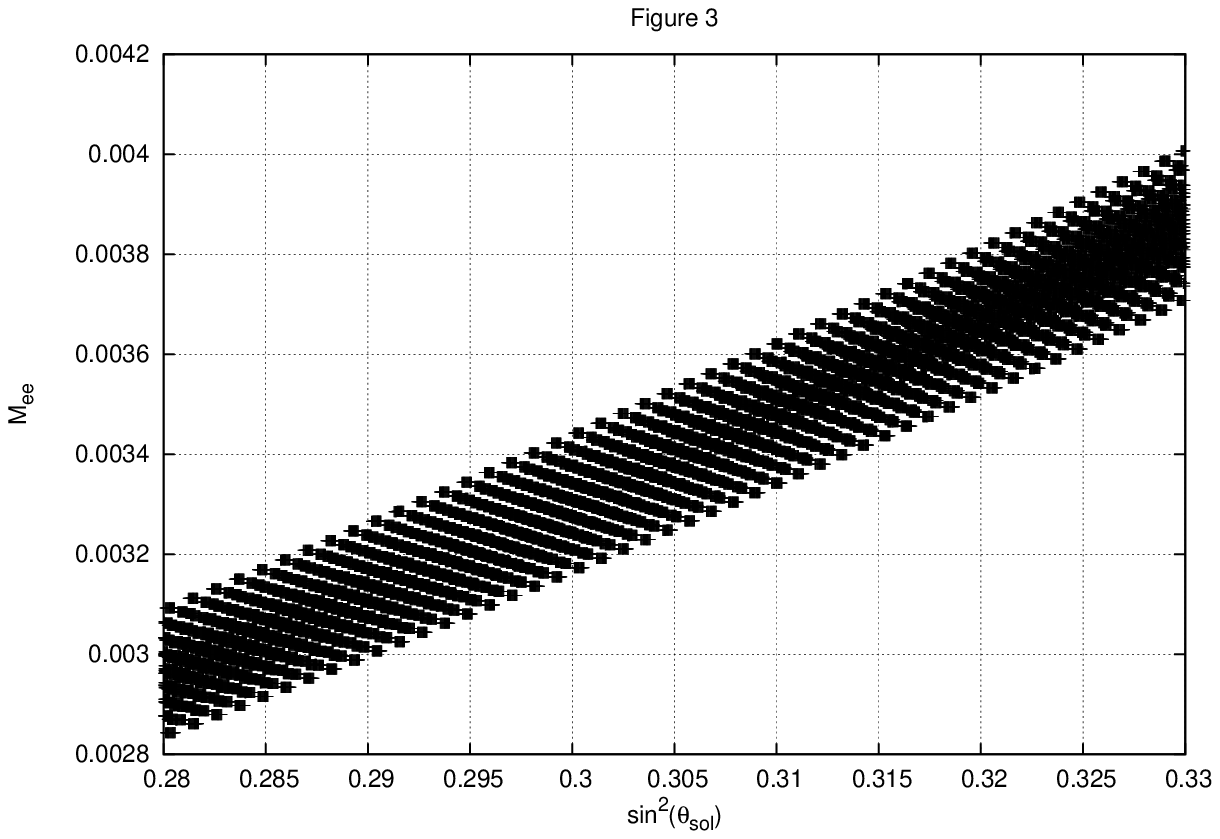}}
\caption{$M_{ee} $ as a function of $\sin^{2}{\protect\theta_{sol}}$,
assuming $1\protect\sigma$ uncertainties in neutrino mass differencies}
\label{fig3}
\end{figure}

\begin{figure}[tbp]
\centerline{\ \epsfysize=8.0truecm \epsffile{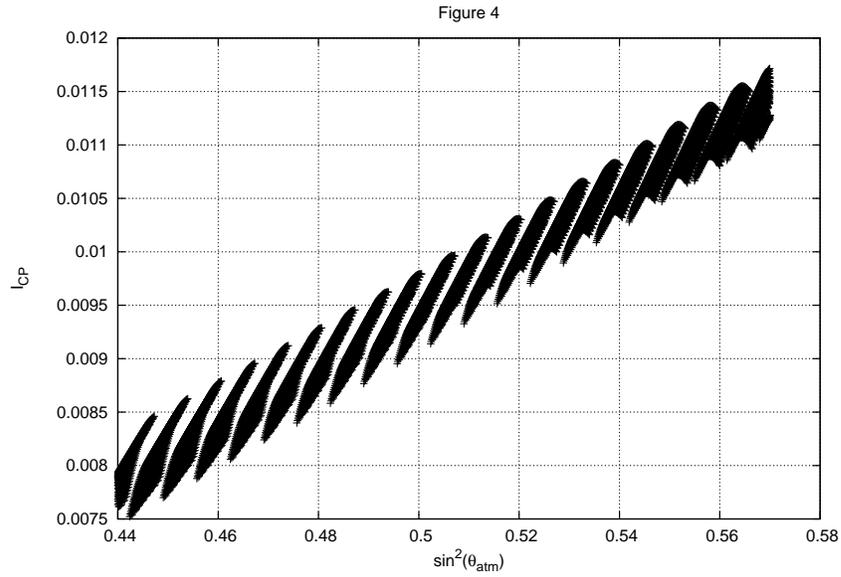}}
\caption{$I_{CP}$ as a function of $\sin^{2}{\protect\theta_{atm}}$,
assuming $1\protect\sigma$ uncertainties in neutrino mass differencies }
\label{fig4}
\end{figure}

\end{document}